\documentclass[aps,prl]{revtex4}
\usepackage{amsfonts}
\def\ee{\end{equation}}
\def\bea{\begin{eqnarray}}

\def\bra#1{\langle #1 |}
\def\ket#1{| #1\rangle}

\def\braopket#1#2#3{\langle \, #1 \, | \, #2 \, | \, #3 \rangle}

\usepackage{graphicx}
\DeclareGraphicsExtensions{.png}

\begin{document}
\title{Perception Constraints on Mass-Dependent Spontaneous Localization}

\author{Adrian \surname{Kent}}
\affiliation{Centre for Quantum Information and Foundations, DAMTP, Centre for
  Mathematical Sciences, University of Cambridge, Wilberforce Road,
  Cambridge, CB3 0WA, U.K.}
\affiliation{Perimeter Institute for Theoretical Physics, 31 Caroline Street North, Waterloo, ON N2L 2Y5, Canada.}
\email{A.P.A.Kent@damtp.cam.ac.uk} 

\date{June 2018 (updated July 2019)} 

\begin{abstract}
Some versions of quantum theory treat wave function collapse as a
fundamental physical phenomenon to be described by explicit
laws. One motivation is to find a consistent unification of quantum
theory and gravity, in which collapse prevents superpositions of
space-times from developing. Another is to invoke collapse to
explain our perception of definite measurement outcomes. Combining
these motivations while avoiding two different collapse postulates
seems to require that perceptibly different physical states
necessarily create significantly different mass distributions in our
organs of perception or brains. 

Bassi, Deckert and Ferialdi investigated this
question in the context of mass density dependent spontaneous
collapse models. By analysing the mechanism of visual perception of
a few photons in the human eye, they argued that collapse model
parameters consistent with known experiment imply that a collapse
would take place in the eye within the human perception time of
$\approx100~$ms, so that a definite state of observing some or no photons
would be created from an initial superposition.  I reanalyse their
arguments, and note a key problem: they treat the relevant processes
as though they take place in vacuo, rather than in cytoplasm. This
makes a significant difference, since the models imply that
superpositions collapse at rates that depend on the difference
between the coarse grained mass densities of their components. This
increases the required collapse rate, most likely by at least an
order of magnitude and plausibly by significantly more. This casts
some doubt on the claim that there are collapse model parameters
consistent with known experiment that imply collapse times of 
$\lesssim$100ms within the human eye. A complete analysis would require a very
detailed understanding of the physical chemistry and biology of rod
cells at microscopic scales.

\end{abstract}
\maketitle
  
\section{Introduction}

Finding a theory that unifies quantum theory and gravity is
universally agreed to be a fundamental unsolved problem in physics. 
Finding a theory that explains the apparent emergence of classicality
from quantum theory, resolving the so-called ``measurement problem''
or ``reality problem'' is thought by many to be another, and 
there are several well-known lines of thought on possible solutions.
Explaining the emergence of consciousness from either classical or
quantum physics is also thought by many to be a fundamental problem;
those who think this mostly think we do not currently have lines of 
thought that promise anything like a complete solution.   

One popular approach to the measurement problem is to propose
explicit laws governing wave function collapse.   Wigner
\cite{wigner1995remarks} considered the possibility that collapses take place when
observations are made by conscious observers.  
Diosi \cite{diosi1987universal} and Penrose \cite{penrose1996gravity}
suggested that unifying quantum theory and
gravity may require that superpositions collapse whenever
they would otherwise create superpositions of
distinguishable spacetimes.  
Ghirardi-Rimini-Weber-Pearle \cite{ghirardi1986unified,ghirardi1990markov} developed spontaneous collapse
models, in which unitary quantum dynamics are replaced by
stochastic differential equations that are proposed as
fundamental laws, from which the unitary Hamiltonian evolution
of micro-systems and the effective collapse of macroscopic 
superpositions emerge as special cases.    
In the currently preferred versions of these models, collapse
rates are proportional to mass densities.   This avoids the need
to treat composite particles such as nucleons as composed of definite numbers of
elementary particles, which would be difficult to reconcile with
current theory.  It also maintains consistency with current
experiments, which appear to exclude the original GRW
\cite{ghirardi1986unified} model.
Moreover, appealingly, it suggests a link with gravity.    

Although all of these justifications are certainly questioned, collapse
hypotheses thus also risk over-motivation.   
It is not immediately obvious that a collapse law designed
to prevent spacetime superpositions necessarily also explains
the appearance of classical outcomes of all measurements,
or even that it is possible to find a single law that does
both and remains consistent with known experiment. 
In principle, of course, one could postulate two or even
more collapse laws: Wigner and Diosi-Penrose could
be pointing to independent fundamental collapse phenomena, for
example.   For most theorists, though, this seems at least one
law too many.  We would like any alternatives to unitary quantum
dynamics to be as simple and elegant as possible and to explain
as much as possible.  

To define and analyse the question quantitatively, we need to 
consider specific dynamical collapse models.   
I focus here on mass-dependent spontaneous collapse models,
and on a pioneering paper \cite{bassi2010breaking}
by Bassi, Deckert and Ferialdi (BDF) which considered
the implications of these models for events associated with
visual perception.    These are certainly not the only
models linking gravity with collapse, and indeed do so
less directly than other proposals.   However, they 
are better developed than most, their experimental
implications have been carefully analysed, and they
include two parameters that allow the predictions 
of other models to be compared and fitted in a given experimental regime.  

On this complex topic, it is natural that some assumptions 
may be debatable, and progress is likely to be incremental.  
Indeed, reanalysing BDF's arguments, I note some
problems both with the calculations and the approximations. 
These make a significant enough difference -- a factor of 
at least $\approx 10$ and perhaps significantly more 
in the lower bound on the collapse rate -- that they cast
doubt on the conclusion that the relevant collapse models 
can be consistent both with known experiment and with 
collapse taking place within human perception times. 

That said, a definitive conclusion would require a very complicated 
analysis, including a detailed understanding of physical chemistry, microscopic cell biology
and the correlates of conscious visual perception in the human
brain.   I am unable to present such an analysis, and indeed
not certain that the present state of understanding of these topics 
will allow precise and reliable estimates of collapse rate bounds 
from perception.   
Nonetheless, more progress can surely be made, and
I hope that this discussion will stimulate further work.

\section{BDF on continuous spontaneous localization}

To ensure that we represent BDF accurately, we quote directly
from their analysis in this and the next section.
BDF begin by presenting the stochastically modified Schr\"odinger 
equation that defines the mass proportional version of the
Ghirardi-Pearle-Rimini \cite{ghirardi1990markov} continuous spontaneous localization
model:
\begin{equation}\label{csleqn}
\ket{d \psi_t} = \left( -\frac{i}{\hbar} H dt + \sqrt{\gamma} \int d^3
  x ( M( {\bf x} ) - \langle M ( {\bf x} ) \rangle_t ) dW_t ( {\bf x}
  )
- \frac{\gamma}{2} \int d^3 x ( M( {\bf x}) - \langle M ( {\bf x} )
\rangle_t )^2 dt \right) \ket{ \psi_t } \, . 
\end{equation}

Here $H$ is the Hamiltonian and $M({\bf x})$ is a smeared mass density
operator.   It takes the form 
\begin{equation}\label{smear}
M({\bf x}) = \frac{1}{m_N} \int d^3 y g ( {\bf x} - {\bf y} ) \sum_s
m_s a^\dagger_s ( {\bf y} ) a_s ( {\bf y } ) , 
\end{equation}
where the sum is over particle species $s$ with mass $m_s$.
BDF take $m_N$ to be the mass of a nucleon, in an approximation
in which the difference between the proton and neutron masses
is negligible.  The smearing function is taken to be
\begin{equation}\label{smearfn}
g ({\bf x}) = \frac{1}{ ( 2 \pi r_C^2 )^{3/2}} \exp ( - {\bf x}^2 / (2
r_C^2 )) \, .  
\end{equation}
Here the coupling constant $\gamma$ and the 
length scale $r_C$ are parameters of the collapse model.
These may be varied independently, and a complete analysis
would consider all ranges of both.   In their analysis BDF
set $r_C \approx 10^{-5}$cm and consider
the bounds implied for $\gamma$, or equivalently for the 
collapse rate
\begin{equation}
\lambda = \frac{\gamma}{8 \pi^{3/2} r_C^3} \, .
\end{equation}

BDF then consider a superposition of states of $N$ particles, 
of the form 
\begin{equation}
\alpha' \ket{{\bf \bar{x}'}} + \alpha'' \ket{{\bf \bar{x}''}} \, , 
\end{equation}
where ${\bf \bar{x}'} = {\bf x'_1 , x'_2 } \ldots {\bf x'_N}$
and ${\bf \bar{x}''}$ is similarly defined. 
(Here BDF implicitly assume that each particle has the nucleon
mass $m_N$: an atom with atomic mass $x$ Daltons is 
effectively treated as a system of $x$ tightly bound nucleons
in their discussion.) 
They set the Hamiltonian to zero, and writing the stochastic
average density matrix as
\begin{equation}
\rho_t = {\mathbb E} \left[ \ket{\psi_t} \bra{\psi_t } \right] \, . 
\end{equation}
They then give the time evolution of the off-diagonal elements:
\begin{equation}
\frac{\partial}{\partial t} \braopket{{\bar{\bf x}}'}{\rho_t}{{\bar{\bf x}}''}
= - \Gamma( {\bar{\bf x}}' , {\bar{\bf x}}'' ) \braopket{{\bar{\bf
      x}}'}{\rho_t}{{\bar{\bf x}}''} .
\end{equation}
Here 
\begin{equation}\label{Geqn}
\Gamma( {\bar{\bf x}}' , {\bar{\bf x}}'' ) = 
\frac{\gamma}{2} \sum_{i,j=1}^N \left[ G( {\bf x}'_i - {\bf x}'_j )  
+  G( {\bf x}''_i - {\bf x}''_j ) 
- 2  G( {\bf x}'_i - {\bf x}''_j ) \right] \, , 
\end{equation}
and 
\begin{equation}
G ({\bf x}) = \frac{1}{ ( 4 \pi r_C^2 )^{3/2}} \exp ( - {\bf x}^2 / (4 
r_C^2 )) \, .  
\end{equation}

Now if $|{\bf x}'_i - {\bf x}''_i | \ll r_C$ for all $i$, then the
first two terms in each summand in Eqn. (\ref{Geqn}) cancel the
third, up to negligible contributions, and so the decay rate
is negligible.   If $|{\bf x}'_i - {\bf x}''_i | \geq 3 r_C$ for
all $i$ while $|{\bf x}'_i - {\bf x}'_j | \ll r_C$ and 
$|{\bf x}''_i - {\bf x}''_j | \ll r_C$ for all distinct
$i,j$, then $\Gamma \approx \gamma ( 4 \pi r_C^2 )^{-3/2}( N^2 - 2N
) $ and so $\Gamma \approx  \gamma ( 4 \pi r_C^2 )^{-3/2} N^2 =
\lambda N^2$ to
leading order in $N$.  If first and second (or third) conditions hold, 
while the third (or second) set of separations are larger than $3 r_C$, then 
$\Gamma \approx  \frac{\gamma}{2} ( 4 \pi r_C^2 )^{-3/2} N^2 =
\frac{\lambda}{2} N^2 $, again
giving a quadratic leading order dependence.
If $|{\bf x}'_i - {\bf x}'_j | \geq 3 r_C$ and
 $|{\bf x}''_i - {\bf x}''_j | \geq 3 r_C$ for all distinct pairs
 $(i,j)$, while $|{\bf x}'_i - {\bf x}''_j | \geq 3 r_C$ for all
$(i,j)$ then only the terms with $i=j$ in the first two sums
contribute, giving $\Gamma \approx \gamma ( 4 \pi r_C^2 )^{-3/2} N =
\lambda N$, i.e. a linear dependence. 

More generally, consider a superposition of two states, in each which
the particles are clustered in groups, with separations $\ll r_C$
within the clusters and $\gg r_C $ between the clusters. 
Suppose that the separations between the states of each cluster
in the two components are $\gg r_C$ and that there are $n_i$ particles in
cluster $i$.  Then to leading order the collapse rate is
\begin{equation}\label{cluster}
\Gamma = \lambda \sum_i n_i^2 \, .
\end{equation}

As noted above, an atom of mass $x$ is treated as a cluster of $x$
nucleons.   As this suggests, one can extend the result to the
general case in which particle type $i$ has mass $m_i$, giving \cite{adler2007lower}
\begin{equation}\label{gencluster}
\Gamma = \frac{\lambda}{m_N^2} \sum_i m_i^2 n_i^2 \, .
\end{equation}

\section{BDF on visual perception} 

BDF consider a human observing a superposition state of a few photons,
arranged so that one component causes the photons to impinge on the
retina while the other does not.   The components of the photon state may be very widely
separated: non-relativistic collapse models generally do not assume
any spontaneous collapse of photon states, and in any case the 
collapse rate for a few particles is negligible and effectively independent 
of the state separation $l$ in the regime $l \gg r_C$. 

The goal of collapse models is to explain the appearance of
classicality.   Humans do indeed perceive definite outcomes -- 
namely, observing photons or not -- when observing such states.  
Hence, BDF argue, a plausible collapse model must imply 
that a superposition reaching the eye must collapse before
it is transformed into a perception in the brain. 
Human reaction time for weak light perceptions is $\approx 100$ms,
so, BDF argue, this requires a collapse within that time. 
This appears reasonable, though of course there is room for
discussion.   Three points seem worth elaborating on. 

First, our reports and memories of perceptions might 
not be entirely reliable.   Theoretically, one could imagine
that collapses take place at a much later point -- hours or
days after the interaction -- leaving us with post-collapse
memory states indistinguishable from memories of a (near)
real time observation.   However, if we are happy to accept
theories in which the appearance of classicality is a 
false post hoc construct, we may struggle to explain 
why we are not happy with some version of many-worlds
quantum theory \cite{saunders2010many}, undercutting entirely the motivation 
for considering collapse models.  

Second, one could imagine that collapse takes place not as
a result of events within the brain, but as a result
of our physiological responses to these events. 
Perhaps neither the eye detecting the photons, nor
our visual cortices processing the information, are
sufficient to cause collapse.   Perhaps, instead,
collapse only takes place when we blink, or subtly shift
position, or report our observation orally or in writing.
Though this is not a ridiculous hypothesis, it is not
completely evident that it is consistent with our 
experience.  
It is tempting, if perhaps naive, to  feel one would surely notice
the photons even if one's head and body were completely immobilized.
Perceptions of conscious events are notoriously tricky and sometimes
deceptive, though (e.g. \cite{libet1985unconscious}). 
Perhaps subtle but macroscopic
involuntary physiological responses could be crucial to 
conscious observation. 

This possibility has been discussed in the past by some advocates of 
CSL \cite{aicardi1991dynamical}.   At present, my impression is that
there is no consensus among advocates of CSL as to how seriously
to take it.  For example, the recent analysis 
Ref. \cite{torovs2018bounds} proposes lower bounds on collapse
model parameters without allowing for the possibility that
physiological responses induce the relevant collapses. 
Everyone should agree, at least, that if a purportedly fundamental physical theory such as a CSL model
can only be kept alive by invoking the hypothesis, 
then (a) anyone advocating the theory should be very clear about 
this, and (b) we should try to test the hypothesis directly as
far as possible (difficult though this is).  
Since my focus here is on BDF's arguments, which do not involve
the hypothesis, I will not consider it further here.    

Third, one could imagine that collapse takes place as a result
of events within the brain, but not necessarily within the eye.
As BDF note, some authors have produced bounds for CSL models
on this hypothesis.   BDF consider it dubious: they argue that
it would imply that ``animals with a simpler visual apparatus
could perceive $\ldots$ superpositions which we consider rather
unlikely''.    Here, if I understand BDF correctly, I disagree.
Animals with simpler brains would not necessarily
perceive superpositions if no collapse took place as
a result of events in their brains before their reaction time.
They might have no conscious perception at all of these
observations, or they might have delayed perceptions.
They might not necessarily have conscious memories of 
these perceptions, and if they do these may not necessarily
give them the same impression of time sequencing that our
memories give us.   So I consider the hypothesis that 
collapse takes place within the human brain, but not necessarily
within the human eye, within $\approx 100$ms, perfectly reasonable.  
However, my aim here is to discuss BDF's arguments.
These assume that
collapse takes place within the eye, and this is certainly
an interesting and prima facie plausible hypothesis.   I will argue that 
there are problems with those arguments, which make it 
very hard to produce precise bounds for mass-dependent CSL collapse
rates.   The same issues arise in considering information 
processing elsewhere in the brain, and so I will not pursue
this hypothesis further here either. 

BDF's account of the biochemical processes 
involved in photodetection in the eye considers
the following stages.  Each photon is absorbed
by a rhodopsin molecule, transforming it.
The transformed molecule interacts with $\approx 20$
transducin molecules, splitting off $\alpha$-subunits
from each.  Each subunit diffuses over the rod disc 
and binds to a phosphodiesterase (PDE) molecule,
activating it.   
Each active PDE converts a cyclic guanosine monophosphate (cGMP) 
molecule to guanosine monophosphate (GMP).   The reduction
in cGMP causes the closure of $\approx 300$ ionic channels
on the rod cell membrane, each preventing $\approx 10$ sodium
ions (${\rm Na}^+$) from entering the rod.  This generates
an electric signal which is transmitted to the optic nerve. 

Using the approximations described in the previous section,
BDF argue that there are three relevant components in
the superposition state of detecting and not detecting 
a photon.   First, the $\approx 20$ $\alpha$-subunits either remain
attached to the transducins or 
diffuse over the rod disc surface, in which case
they become separated from one another 
by $> r_C$.  They then bind to PDE.    
Second, in the absence of photons cGMP molecules bind
to the ion channels, while converted GMP molecules diffuse
in the cytoplasm. 
Third, $\approx 10^3$ ${\rm Na}^+$ ions either enter or fail to 
enter the rod membrane through ion channels.   

BDF argue that Eqn. (\ref{cluster}) can be applied to obtain
contributions to the collapse rate from each of these three
components.  They take the first component as effectively 
giving a contribution of $n_1^2 N_1$, where $ n_1 = 3.9 \times 10^4$ 
is the molecular weight of the $\alpha$-subunits in daltons,
and $N_1 = 20$ is the number of subunits separated by $>r_c$.
The second component is taken to give a contribution 
$n_2^2 N_2$, where $n_2 = 363$ is the molecular weight of GMP
and $N_2 = 2000$ the number of molecules. 

The third component is taken to give a contribution 
$n_3^2 N_3$, with two different hypotheses assigning
different values.  One (BDF's ``most likely case'')
takes $n_3 = 5 \times 3 \times 23$, corresponding
to $5$ channels within distance $r_C$, clusters of 
$3$ ions separated by $<r_C$, each with molecular weight
$23$.   In this case there are $\approx 60$ groups
of $5$ channels, and $\approx 333$ clusters of ions,
and BDF take $N_3 = 60\times 333$. 
The second (BDF's ``extreme case'') assumes all
ions passing through a channel are separated 
by $<r_C$, giving them $10^3$ ions for each of $5$ channels 
in a cluster and $n_3 = 5 \times 10^3 \times 23$, and
$60$ groups of $5$ channels, giving $N_3 = 60$.  

Accepting these values for the moment, this gives
\begin{equation}\label{bdfrangeone}
n_1^2 N_1 \approx 3 \times 10^{10} \, , \qquad
n_2^2 N_2 \approx 3 \times 10^8  \, , \qquad 
\end{equation}
and two estimates defining a range for the third
contribution
\begin{equation}\label{bdfrangethree}
n_3^2 N_3 \approx 2 \times 10^9 - 8 \times 10^{11} \, . 
\end{equation} 
Summing these, BDF argue, gives the effects of one photon,
and multiplying by $6$ gives the effects of $6$ photons,
which they take to be the fewest detectable by the human
eye.   Thus their final estimate is 
\begin{equation}
6 \times \sum_{i=1}^3 n_i^2 N_i \, , 
\end{equation}
with the $n_i$ and $N_i$ given above.   

\section{Problems in the BDF analysis}

\subsection{Problems in BDF's calculations}

The second term in BDF's sum is dominated by
the first and third, and so may be neglected.
The first term lies within the range of estimates
for the third, so that the sum lies in a compressed
range
\begin{equation}
6 \times \sum_{i=1}^3 n_i^2 N_i \approx 2 \times 10^{11} - 5 \times
10^{12 } \, .
\end{equation}
BDF's estimates for $\lambda$ appear, however, to be based
only on their estimates for the range of $n_3^2 N_3$, 
neglecting the contribution of $n_1^2 N_1$.   
This gives them a larger range than should follow from
their assumptions and estimates.   

BDF adopt the criterion that a superposition is taken to have collapsed 
when $\Gamma t \approx 10^2$, meaning that one term is $\approx
e^{100}$ times smaller than the other.   As they note, this is 
reasonable but arbitrary, and a factor of $10$ either way could 
reasonably be included.   
The equation $\Gamma t = 10^2$, with a time $t = 100$ms,
implies $\Gamma = 10^3 {\rm s}^{-1}$. 

Using the corrected range, we find from BDF's estimates 
and Eqn. (\ref{cluster}) a range for the collapse rate given
by
\begin{equation}
\lambda \approx 5 \times 10^{-9} - 2 \times 10^{-10} \, , 
\end{equation}
rather than BDF's estimate of 
\begin{equation}
\lambda \approx 5 \times 10^{-9} - 2 \times 10^{-11} \, , 
\end{equation}
As BDF note, both ranges could reasonably be multiplied 
by $10^{\pm 1}$ given the arbitariness noted in the previous
paragraph.   

\subsection{Allowing for the cytoplasm}

BDF's calculations effectively model visual perception as
though the only relevant massive particles are the 
specific particles they discuss: the $\alpha$-subunits,
the GMP molecules, and the ${\rm Na}^+$ ions.   Their
estimates of the collapse rate are thus derived from
Eqns. (\ref{Geqn}) and (\ref{cluster}), where the sums
include these particles and no others.    

This would be a valid approximation if the interactions 
between incoming photons and these three types of particles
took place in otherwise empty space.   In fact, of course,
they take place within rod cells, which have membranes and
other structures filled with cytoplasm, a gel-like substance
containing many proteins and ions.   

To see immediately that this is likely to affect the calculations 
significantly, note that Eqn. (\ref{csleqn}) depends on 
$( M( {\bf x} )$ through $( M( {\bf x} ) - \langle M ( {\bf x} )
\rangle_t )$, and that $M({\bf x})$ itself is a smeared mass
density, with the smearing function (\ref{smearfn}) having
characteristic scale $r_C$.   

\subsubsection{Considering the cytoplasm as homogeneous}

We thus cannot apply Eqn. (\ref{gencluster}) directly,
taking $m_i$ as the actual masses for the relevant particles, 
for superpositions arising in an otherwise
homogeneous fluid. 
A more relevant approximation would be to take 
\begin{equation}
m'_i = m_i - \rho  V \, , 
\end{equation}
where $m_i$ is the actual particle mass, $\rho$ the 
average smeared density of the fluid, and $V$ the volume
of fluid notionally displaced by the particle.  
More precisely, we could take $\rho V = k_i m$, where $k_i$ is the 
(not necessarily integer) average number of fluid particles 
absent in a volume of $r_C^3$ when that volume contains a 
particle of type $i$ and $m$ is the mass of each fluid particle. 

This is significant because the densities of the relevant particles
in BDF's analysis and of the cytosol and other components of the
cytoplasm are likely not dissimilar.   It is hard to be precise, 
because the details depend on the properties of the relevant 
particles when suspended in the cytosol environment, which
itself is complex.   I have found it hard to locate data 
even for aqueous suspensions.   The best I can offer 
are very crude estimates, which nonetheless illustrate the 
problem and the need for closer analysis.    

For example, the density
of metallic sodium, $968~ {\rm kg}{\rm m}^{-3}$, is very close to that
of water, $997 ~{\rm kg}{\rm m}^{-3}$.   While data on the effective density
of ${\rm Na}^+$ ions in water solution is harder to find, one crude
estimate is given by comparing the estimated effective radius of ${\rm
  Na}^+$ in water \cite{yang2015size} ($218~$pm), by that of ${\rm Na}$ atoms
($227~$pm).  If (which is admittedly not clearly justified by the cited data) 
we could approximate the effective density
of ${\rm Na}^+$ in water by $ (227/219)^3 \, 968 \approx 1078~ {\rm kg}{\rm
  m}^{-3}$, we would get an effective $m'_i$ for sodium ions in water
of approximately $ 0.08 \, m_i$, thus multiplying the estimated collapse rate
by $< 10^{-2}$.   

To get a crude estimate for the $\alpha$-subunits and GMP molecules,
we could compare the typical density of proteins, $\approx 1200-1400{\rm kg}{\rm
  m}^{-3}$, with either the density of water or, presumably better,
the density of the rod cytosol or cytoplasm (perhaps $1100 {\rm kg}{\rm
  m}^{-3}$).   This gives an effective $m'_i \approx 0.3 m_i$,
multiplying
the estimated collapse rate in the rod by $\approx 10^{-1}$ in this
case.   

Allowing for these factors gives a collapse rate estimate in the rod of
\begin{equation}
6 \times \sum_{i=1}^3 (m'_i )^2 n_i^2 N_i \approx 2 \times 10^{10} - 5 \times
10^{10} \, .
\end{equation}

This would imply bounds in the range
\begin{equation}\label{rangetwo}
\lambda \approx 2 \times 10^{-8} - 5 \times 10^{-8} \, . 
\end{equation}
\vfill

\pagebreak

\begin{figure}[h]
\centering
\includegraphics[width=0.5\linewidth, height=5cm]{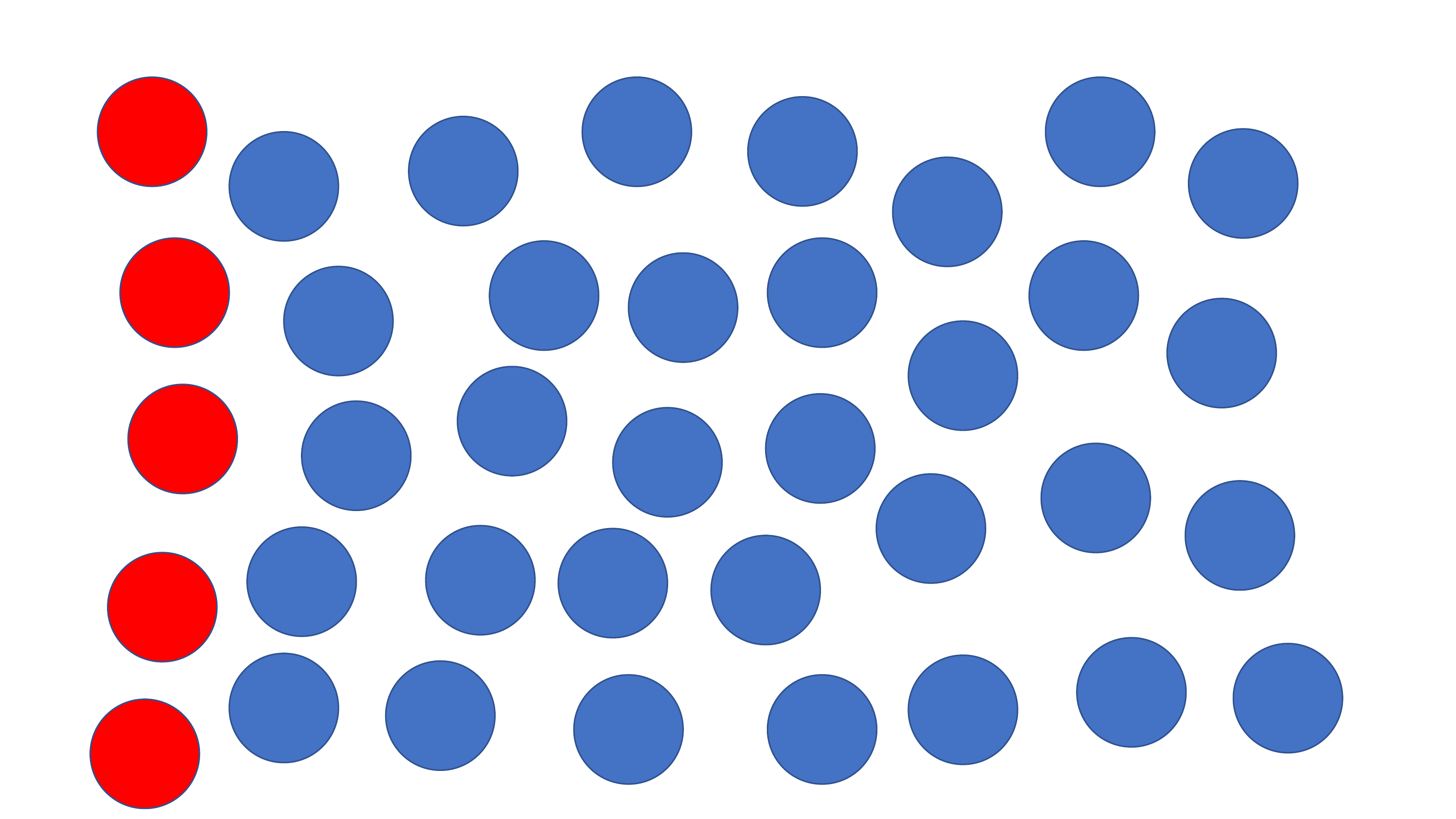} 
\caption{Ions concentrated on left}
\label{ionconc}
\end{figure}

\begin{figure}[h]
\includegraphics[width=0.5\linewidth, height=5cm]{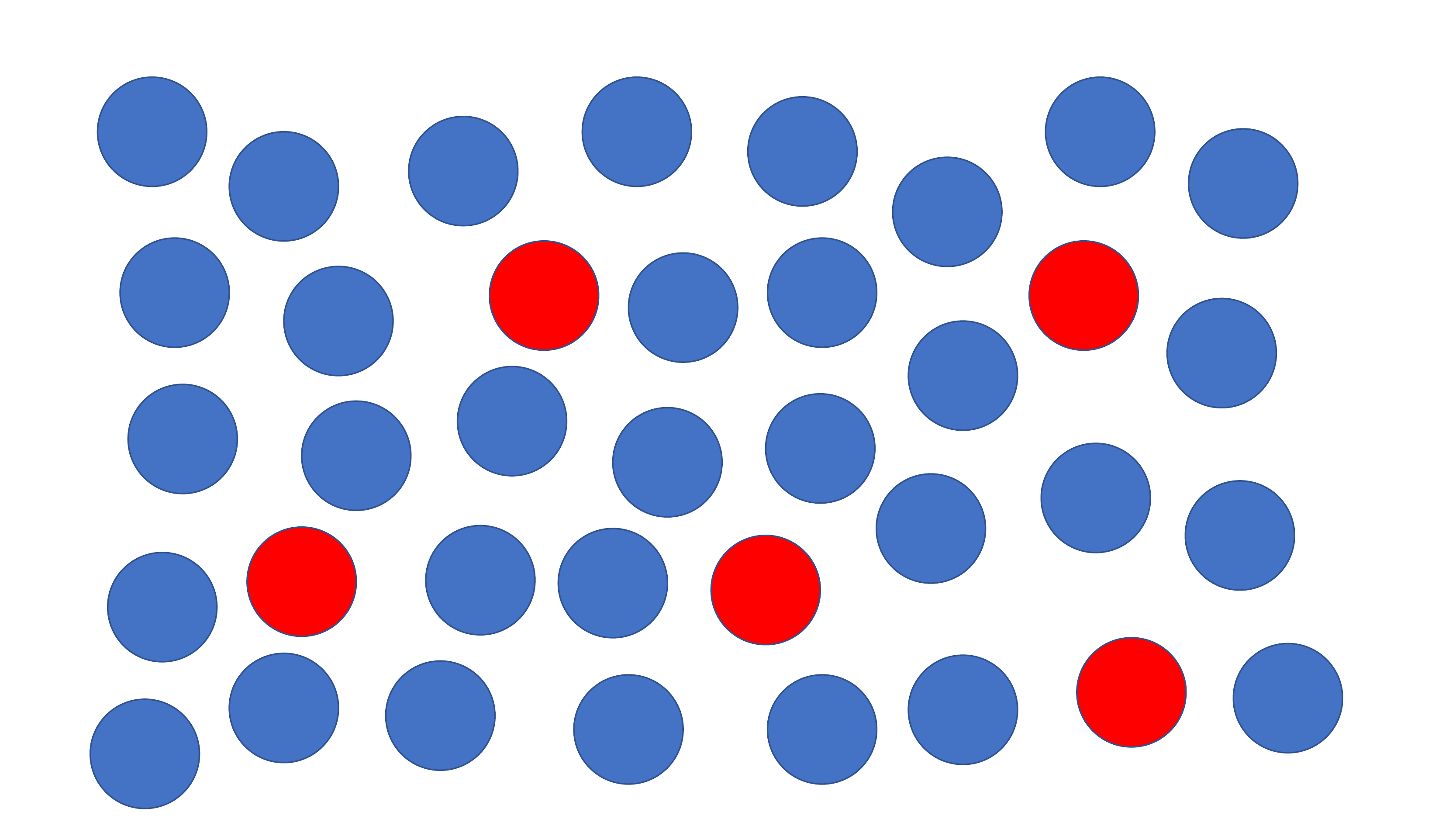}
\caption{Ions diffused}
\label{iondiff}
\end{figure}

Figures \ref{ionconc} and \ref{iondiff} give schematic illustrations of superposition states
illustrating the relevance of relative densities.
Here the red dots represent idealized ions and the
blue dots idealized fluid molecules.   
In the first state, the ions are concentrated at
the edge of the volume; in the second, they have
diffused throughout the fluid.   
In our simplified model, the ions have the same
mass and volume as the fluid molecules and diffuse
so that the molecule positions are identical (although
different molecule types occupy some positions) in the
two components.
An approximation which considers only the ion positions
would suggest that the two states are significantly distinct,
and hence that the mass-dependent CSL model should predict
the superposition will collapse. 
However, when all the particles are taken into
account, the two states have identical mass
distributions.  Hence the mass-dependent 
CSL model predicts no collapse. 


\subsubsection{Allowing for cytoplasmic inhomogeneity}

Even these last estimates, however, are based on an invalid 
model.   Cytosol and cytoplasm are not at all homogeneous on the
relevant scales.   To calculate the difference in smeared mass density
distributions between a superposition component in which some number
of proteins have or have not diffused around the cell, for example,
one thus has to consider all the proteins and other components that
may have been relocated in the course of the diffusion. 
To then apply (\ref{gencluster}), one needs to know -- or at least
plausibly estimate -- all the relevant separations and displacements  
of all these proteins (including but not only those actively involved
in photo-detection), and all the ions and other solutes.

Without a very detailed understanding of rod cell biology and 
biochemistry at very small scales, it is hard to know how to begin making
a plausible estimate.   Cells appear to be crowded enough by 
proteins of various shapes and sizes that diffusion 
processes for any given protein cannot be well modelled by 
treating the cytosol as a dilute solution of that protein \cite{ando2010crowding}. 

\begin{figure}[h]
\centering
\includegraphics[width=0.5\linewidth, height=5cm]{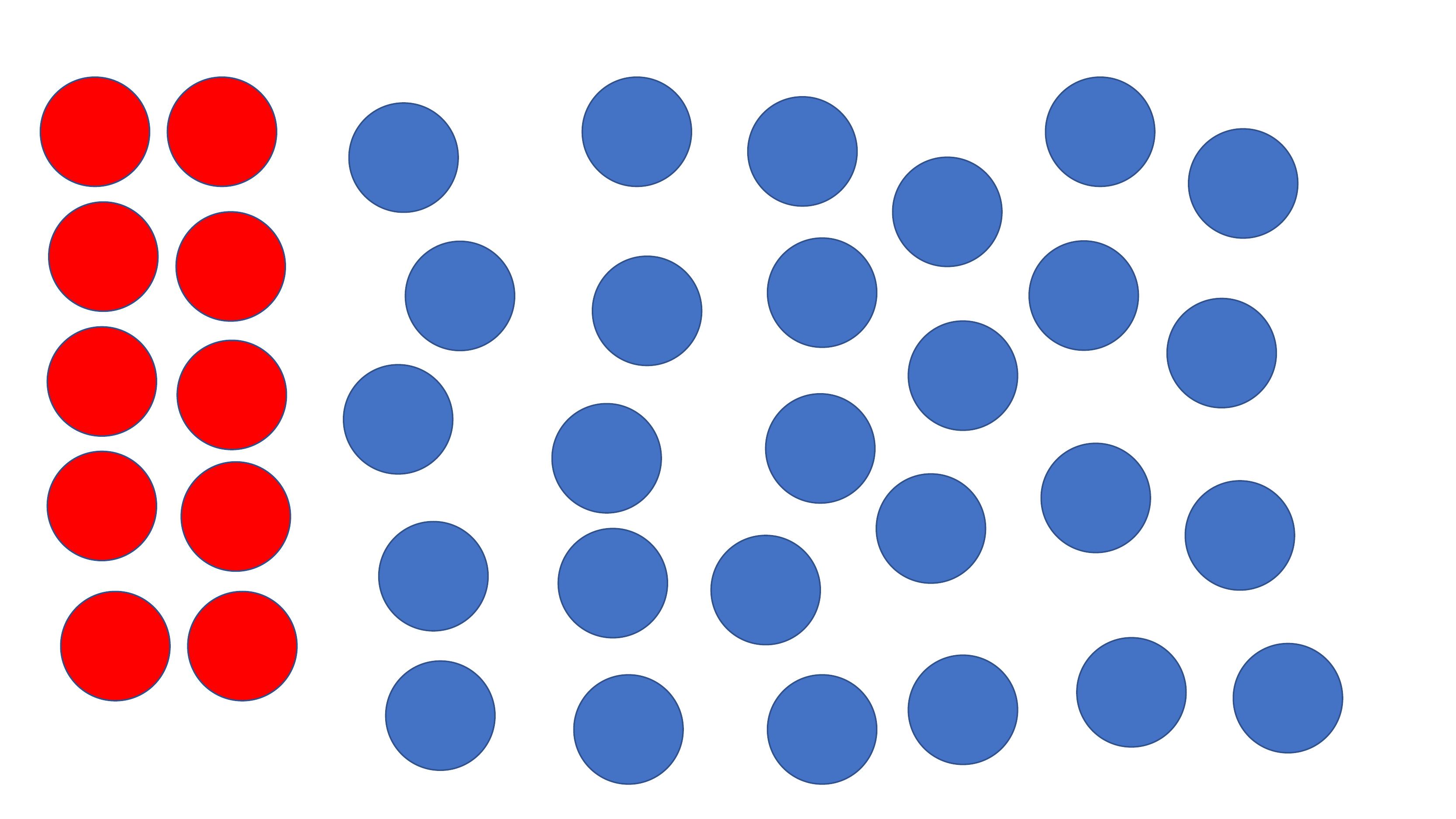} 
\caption{Red protein molecules concentrated on left}
\label{protconc}
\end{figure}

\begin{figure}[h]
\includegraphics[width=0.5\linewidth, height=5cm]{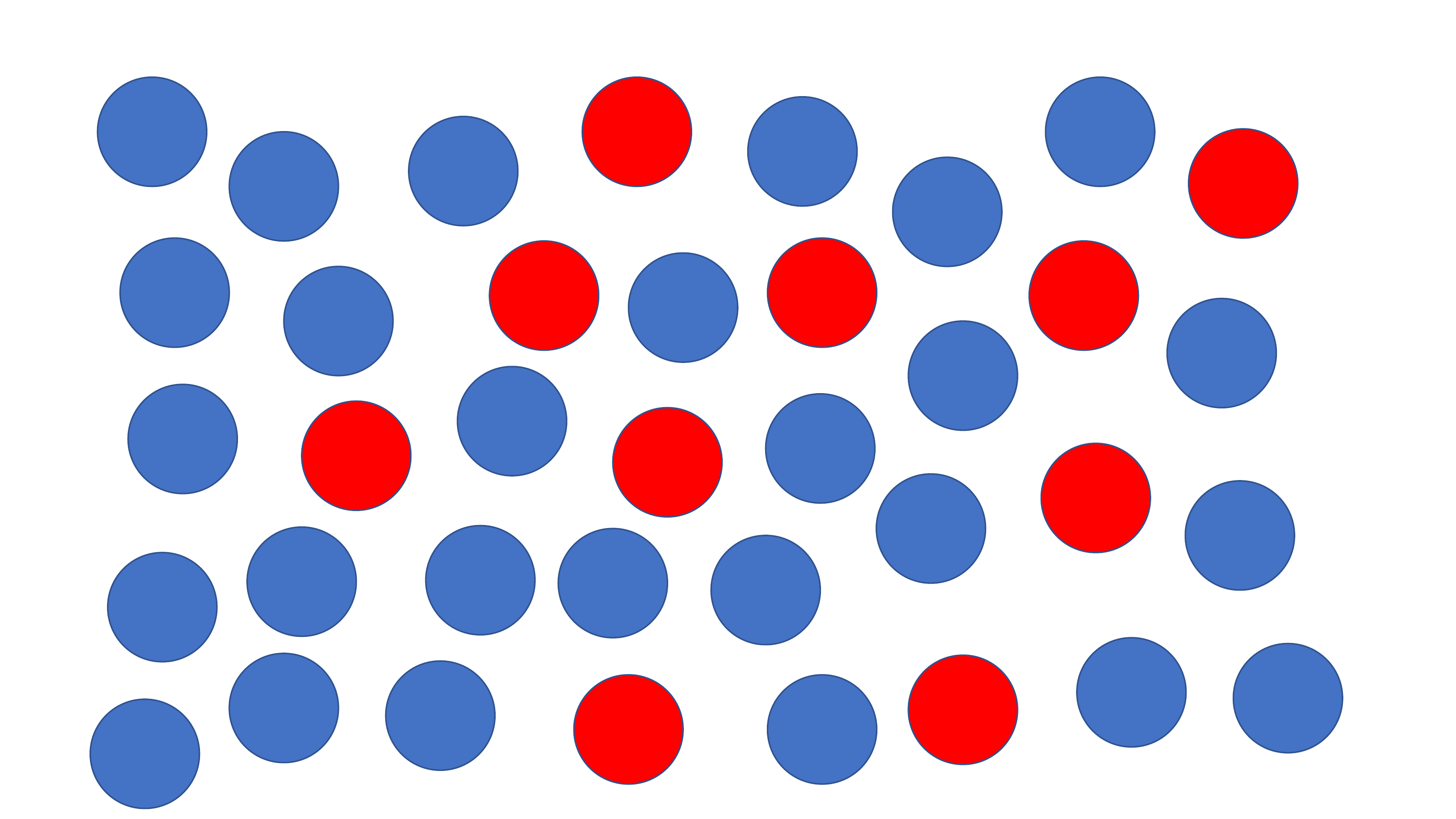}
\caption{Red protein molecules diffused}
\label{protdiff}
\end{figure}

Figures \ref{protconc} and \ref{protdiff} give schematic illustrations of superposition states
illustrating the relevance of inhomogeneities.
Here the red dots represent protein molecules
relevant to visual perception and the
blue dots other protein molecules in the cytoplasm.   
In the first state, the red molecules are concentrated at
the edge of the volume; in the second, they have
diffused throughout the cytoplasm.   
An approximation which considers only the red molecule positions
suggests that the two states are significantly distinct,
and that the separations relevant to the two red
molecule states are large.  
In this model, the molecule positions are different in the
two components.   Thus, 
when all the particles are taken into
account, the two states still have distinct mass
distributions.  However, if the red and blue protein molecules
have identical masses and densities, the relevant separations are
those between dots of either colour in the two component states.
These are much smaller than the typical differences between red
molecule positions in the two states or the typical separations
between red molecule positions in the second state.  


More realistic illustrations can be found in Fig. $1$ and in the
supplementary information of Ref. \cite{ando2010crowding}.  
These suggest a very complex picture of intra-cellular protein
diffusion.   The interaction of any given diffusion process
with collapse dynamics may not be easily captured without
very precise information about the relevant cell environment.   
It is thus impossible to say for sure, but to me the most plausible guess
is that an accurate estimate would produce significantly
higher collapse rate bounds than those derived from Eqn. (\ref{bdfrangeone}). 

Similar comments apply to the collapse rate bounds derived from
the diffusion of sodium ions.   Modelling the sodium ions as 
inhomogeneities in an otherwise homogeneous aqueous solution,
as above, is likely misleading.   One needs to consider the 
precise environment within the solution, including all ions
and other solutes, and allowing for the smearing defined by
Eqn. (\ref{smearfn}).    Again, I am not sure of the likely
result, but find it plausible that the result would be significantly higher collapse
rate bounds than those given in Eqn. (\ref{rangetwo}). 

\subsection{Limits of human perception}

Since BDF's work, evidence has been presented \cite{tinsley2016direct}
suggesting that humans are able to detect single photons.
The evidence is not as yet compelling: results are
reported for three individuals, and their responses
were statistically significant but not perfectly
reliable. 

If it could be shown that humans can reliably detect
single photons, BDF's collapse rate bounds, and
others similarly derived, would be increased by
a further factor of $6$.   Given the uncertainty in 
interpreting the evidence, I do not include
this additional factor here.  It is worth keeping in
mind, though, given that it would increase the bounds
by close to a further order of magnitude. 

\section{Conclusions} 

Dynamical collapse models in general, and mass-dependent continuous
spontaneous localization models in particular, are well motivated
and experimentally testable alternatives to quantum mechanics. 
It is an intriguing question whether these models can be excluded
with forseeable technology, or even are already excluded by 
existing experimental and observational data.   
Lower bounds on the model collapse rates can only ultimately
be justified by assuming that collapses take place within
human perception times, so that the models predict that humans
should perceive one component of a superposition. 

BDF's pioneering work gives a basis for deriving such bounds.
However, their assumptions and approximations are questionable
enough that it seems unwise to rely on the bounds they suggest. 
Further detailed work is needed to decide whether mass-dependent
continuous spontaneous localization models remain viable
(for some parameter choices) or are already effectively
excluded.

\section{Acknowledgements}
This work was partially supported by Perimeter Institute
for Theoretical Physics. Research at Perimeter Institute is supported
by the Government of Canada through Industry Canada and by the
Province of Ontario through the Ministry of Research and Innovation.
I thank Angelo Bassi and Philip Pearle for very helpful discussions.  

\section*{References}

\bibliographystyle{unsrtnat}
\bibliography{braincollapse}{}
\end{document}